\documentclass{nature}

\usepackage{color} 
\usepackage{nicefrac}
\usepackage{amsmath}
\usepackage{amssymb}
\usepackage[hypertexnames=false]{hyperref}
\usepackage{graphicx}
\usepackage[separate-uncertainty = true]{siunitx} 
\usepackage{caption}
\usepackage{sectsty} 
\usepackage{float} 

\usepackage{booktabs}
\usepackage{siunitx} 

\allsectionsfont{\sffamily} 

\usepackage{braket}

\usepackage{lipsum} 

\DeclareCaptionLabelSeparator{bar}{ \textbf{$\vert$} }
\definecolor{stateblue}{rgb}{0.18,0.5,.75} 
\definecolor{statered}{rgb}{0.88,0.29,.29}

\newcommand{\TN}{$T_{\rm N}$ }

\newcommand{\RT}{room temperature }

\newcommand{\FeTe}{Fe$_{1+y}$Te }

\newcommand{\tetra}{$\mathcal{T}$ }

\newcommand{\mono}{$\mathcal{M}$ }

\title{Domain imaging across the magneto-structural phase transition in Fe$_{1+y}$Te}

\author{Jonas Warmuth$^1$, Martin Bremholm$^2$, Philip Hofmann$^3$, Jens Wiebe$^1$, and Roland Wiesendanger$^1$}

\begin{document}

\maketitle

\begin{affiliations}
 \item Department of Physics, Hamburg University, 20355 Hamburg, Germany
 \item Department of Chemistry, Aarhus University, Aarhus, Denmark
 \item Department of Physics and Astronomy, Interdisciplinary Nanoscience Center (iNANO), Aarhus University, 8000 Aarhus C, Denmark
\end{affiliations}

\begin{abstract}
The investigation of the magnetic phase transitions in the parent compounds of Fe-based superconductors is regarded essential for an understanding of the pairing mechanism in the related superconducting compounds\cite{Paglione2010a,Stewart2011,Hoffman2011,Dai2015}. Even though the chemical and electronic properties of these materials are often strongly inhomogeneous on a nanometer length scale\cite{He2011,Singh2013,Kamlapure2017}, studies of the magnetic phase transitions using spatially resolved experimental techniques are still scarce\cite{Enayat2014a,Manna2017}. Here, we present a real space spin-resolved scanning tunneling microscopy investigation of the surface of Fe$_{1+y}$Te single crystals with different excess Fe content, $y$, which are continuously driven through the magnetic phase transition. For Fe$_{1.08}$Te, the transition into the low-temperature monoclinic commensurate antiferromagnetic phase\cite{Hanke2017} is accompanied by the sudden emergence of ordering into four rotational domains with different orientations of the monoclinic lattice and of the antiferromagnetic order, showing how structural and magnetic order are intertwined. In the low-temperature phase of Fe$_{1.12}$Te one type of the domain boundaries disappears, and the transition into the paramagnetic phase gets rather broad, which is assigned to the formation of a mixture of orthorhombic and monoclinic phases\cite{Koz2013}.

\end{abstract}

\FeTe is the non-superconducting parent compound of Fe$_{1+y}$Se$_x$Te$_{1-x}$, in which superconductivity is induced by the substitution of Te with Se \cite{Mizuguchi2010}. These Fe-chalcogenides form the structurally simplest material group of all Fe-based superconductors and they are therefore the ideal compound for a fundamental investigation of the complex mechanisms leading to superconductivity in these materials \cite{Paglione2010a,Stewart2011,Dai2015,Koz2013,Subedi2008}. So far, the complex interplay of magnetic and structural phase transitions in Fe$_{1+y}$Te, which crucially depends on the amount $y$ of excess Fe, has been mostly investigated by spatially averaging techniques such as neutron diffraction, magnetic susceptibility, and resistivity measurements\cite{Koz2013,Zaliznyak2012,Dai2015}. For low values of excess Fe, $y<0.11$, a simultaneous magnetic and structural, so called magneto-structural, transition from a high temperature paramagnetic tetragonal to a low temperature monoclinic phase with a diagonal double-stripe antiferromagnetic spin structure\cite{Hanke2017} was observed. In the following, we will refer to the former phase as the \tetra phase and to the latter one as the \mono phase. The transition temperature of this phase transition decreases with increasing $y$. For $y>0.13$ there is a magneto-structural transition from the \tetra phase into an incommensurate antiferromagnetic orthorombic phase upon cooling\cite{Bao2009b,Rodriguez2011}. Most notably, for $0.12\leq y\leq 0.13$, it was suggested that, upon cooling, the system first transforms from the \tetra phase into an intermediate incommensurate antiferromagnetic orthorombic phase and consecutively into a mixture of the incommensurate antiferromagnetic orthorombic and the \mono phase, suggesting a tricritical point at $y=0.11$\cite{Koz2013}. However, a detailed understanding of the proposed mixed phase and its evolution on a microscopic level were so far lacking.

Former studies using spatially resolved techniques focused on the investigation of the magnetic components of the \tetra to \mono transition in the regime $y<0.11$ as well as on the incommensurate antiferromagnetic orthorombic phase in the regime $y>0.13$ with spin-resolved scanning tunneling microscopy (SP-STM)\cite{Enayat2014a}. For the \mono phase, a twinning of the diagonal double-stripe antiferromagnetic spin structure into two domains rotated by \SI{90}{\degree} was revealed. For $y=0.15$, the SP-STM images indicated coexistence of diagonal double-stripes rotated by \SI{90}{\degree}. To the best of our knowledge, neither the structural components of the \tetra to \mono magneto-structural phase transition in the regime $y<0.11$, nor the complex consecutive phase transitions in the regime $0.12\leq y\leq 0.13$ have been studied on a local scale. Here, we present the first report on continuous temperature-dependent STM measurements mapping the magneto-structural phase transition on \FeTe in real space. We investigated Fe$_{1+y}$Te samples with two different excess Fe contents, $y=0.08$ and $y=0.12$, determined from single crystal X-ray diffraction (SC-XRD, see Methods section), such that the former is expected to have a low temperature \mono phase, while the latter should show the proposed mixed phase\cite{Koz2013}. The experimental methods used for the sample preparation, characterization and the used experimental techniques of magnetic susceptibility measurements and SP-STM are described in the Methods section.

\begin{figure}[H]
      \includegraphics[width = 0.54\columnwidth]{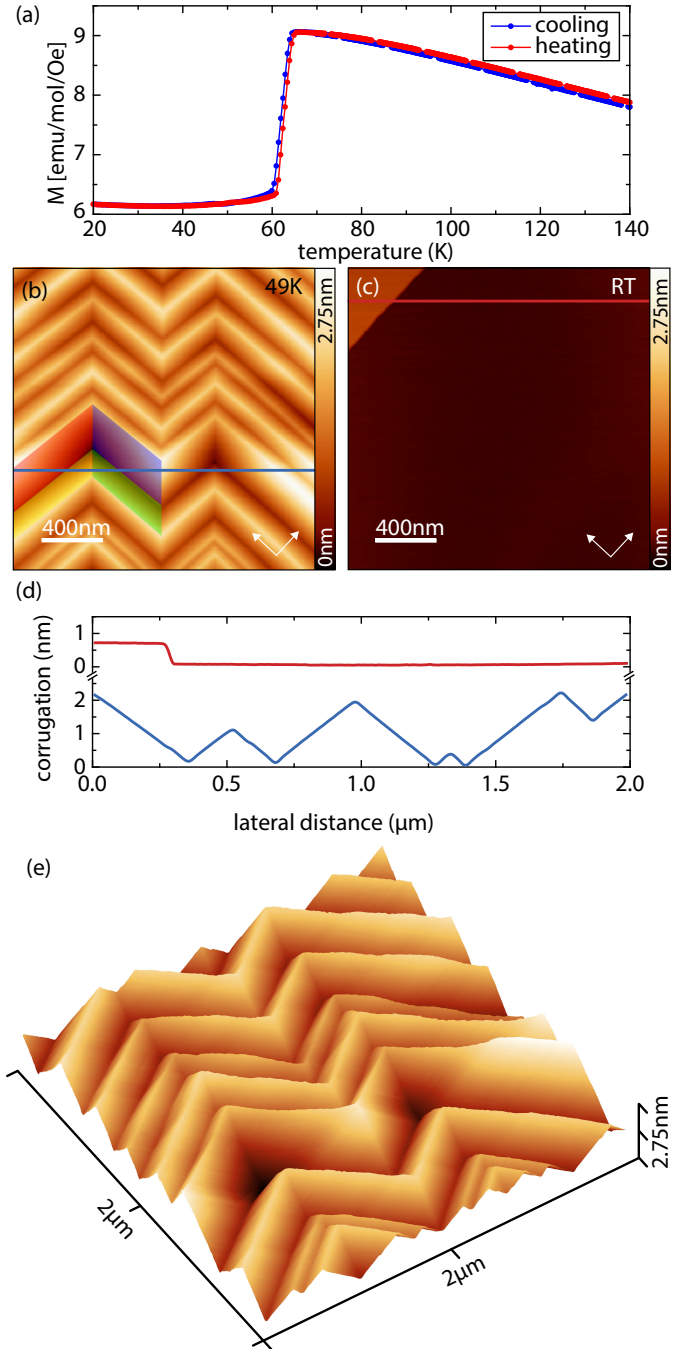}%
      \caption{\label{fig1} \textbf{Surface morphologies of the \mono and \tetra phases of Fe$_{1.08}$Te.} \textbf{a}, Magnetic susceptibility of Fe$_{1.08}$Te measured in a field of \SI{0.1}{\tesla} as a function of temperature for the cooling (blue) and heating (red) cycle. The magneto-structural phase transition appears at \TN$\approx\SI{65}{\kelvin}$. \textbf{b}, Constant-current STM image of the surface of Fe$_{1.08}$Te taken at $T=\SI{49}{\kelvin}$ in the \mono phase ($V_{\rm s}=\SI{100}{\milli\volt}$, $I_{\rm t}=\SI{40}{\pico\ampere}$). Four domains of different types are marked by color. \textbf{c}, Constant-current STM image of the surface of Fe$_{1.08}$Te taken at \RT in the \tetra phase ($V_{\rm s}=\SI{500}{\milli\volt}$, $I_{\rm t}=\SI{20}{\pico\ampere}$). \textbf{d}, Line profiles taken along the lines marked in \textbf{b} and \textbf{c}. \textbf{e}, Three-dimensional view of the surface of Fe$_{1.08}$Te shown in \textbf{b}.}
 \end{figure}

The temperature dependence of the magnetic susceptibility of the sample with the lower content of excess Fe (Fe$_{1.08}$Te) reveals a sharp transition at \TN$\approx\SI{65}{\kelvin}$ with only a small thermal hysteresis ($\Delta T \approx \SI{1}{\kelvin}$) between the cooling and heating cycles (\autoref{fig1}a). As shown in Ref.\cite{Koz2013} and corroborated by the shape of the magnetic susceptibility, the relatively low Fe content is expected to result in a simultaneous first-order magnetic and structural transition from the \tetra into the \mono phase upon cooling. In order to image this transition in real space, the surface of the Fe$_{1.08}$Te single crystal is investigated with temperature dependent SP-STM. In the high temperature \tetra phase (\autoref{fig1}c) the surface is atomically flat apart from step edges with a height of an FeTe layer with a thickness of one unit cell (see upper left corner of \autoref{fig1}c and line profile in \autoref{fig1}d). In strong contrast, in the low temperature \mono phase, the surface decomposes into diamond-shaped domains of hundreds of nanometer length and width, which form a chevron pattern (\autoref{fig1}b and e). Within the domains, the surface is still atomically flat, but the surface normal vectors have four different orientations corresponding to four different domain types classified by colors in \autoref{fig1}b. The surface normal of the green, yellow, blue and red domain types are tilted by $0.81^\circ\pm0.04^\circ$ with respect to the overall surface normal towards the four cardinal crystallographic directions of the Fe$_{1.08}$Te single crystal. This leads to an increased roughness of the surface as seen in the line profile in \autoref{fig1}d.

\begin{figure}[H]
	\includegraphics[width = 0.7\columnwidth]{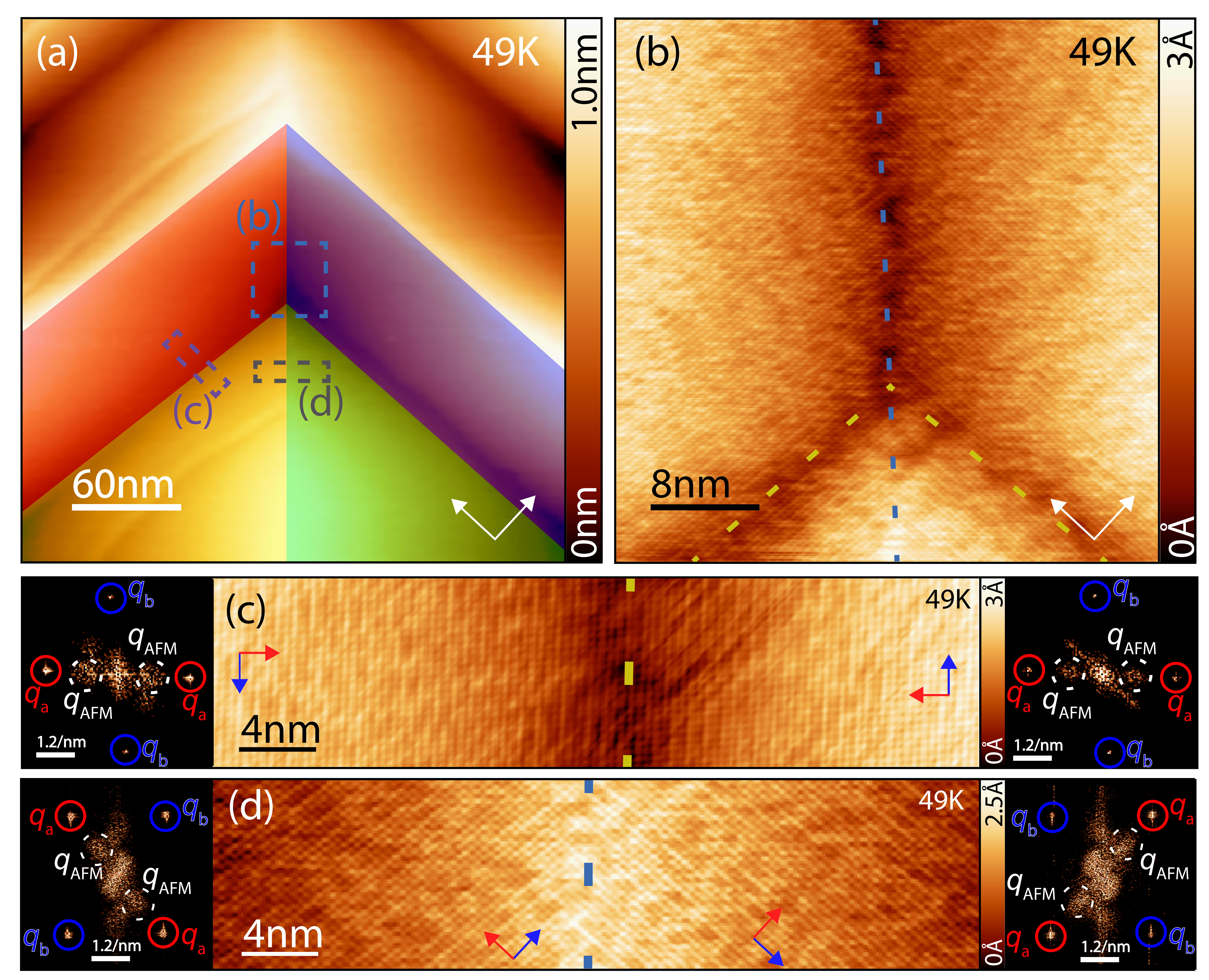} 
 	\caption{\label{fig2} \textbf{Magnetic imaging of the structural domains of the \mono phase of Fe$_{1.08}$Te.} All SP-STM images were recorded at $T=\SI{49}{\kelvin}$. \textbf{a}, Overview constant-current STM image of an area of the \mono phase of Fe$_{1.08}$Te with four domains of different types marked by different colors. The dashed boxes indicate the areas where the images \textbf{b} to \textbf{d} were taken ($V_{\rm s}=\SI{300}{\milli\volt}$, $I_{\rm t}=\SI{40}{\pico\ampere}$). \textbf{b}, SP-STM image of the area marked in \textbf{b} showing the intersection of four structural domains resulting in two A-type (dashed yellow lines) and two B-type (dashed blue lines) structural domain boundaries ($V_{\rm s}=\SI{-100}{\milli\volt}$, $I_{\rm t}=\SI{100}{\pico\ampere}$). The arrows in \textbf{a} and \textbf{b} indicate the lattice directions of the unit cell. \textbf{c}, SP-STM image of the area marked in \textbf{a} with an A-type boundary (yellow dashed line) between a red (left side) and a yellow (righ side) domain type ($V_{\rm s}=\SI{-100}{\milli\volt}$, $I_{\rm t}=\SI{106}{\pico\ampere}$). Note, that the image is rotated by $45^\circ$ with respect to \textbf{a}, \textbf{b} and \textbf{d}. \textbf{d}, SP-STM image of the area marked in \textbf{a} with a B-type boundary (blue dashed line) between a yellow (left side) and a green (right side) domain type ($V_{\rm s}=\SI{-100}{\milli\volt}$, $I_{\rm t}=\SI{102}{\pico\ampere}$). The red and blue arrows in \textbf{c} and \textbf{d} indicate the directions of the $a$ and $b$ lattice vectors, respectively. The insets in \textbf{c} and \textbf{d} represent the FFTs of the left and right sides of the images. Red, blue, and dashed white circles surround the Bragg peaks $q_a$ and $q_b$ and the peaks $q_{\rm AFM}$ due to the diagonal double-stripe spin structure, respectively.}
 \end{figure}

In order to study the magnetic order within these domains, the surface is investigated by atomically resolved SP-STM across the domain boundaries (\autoref{fig2}a,b). As shown in \autoref{fig2}b, as well as in the zoomed images in \autoref{fig2}c,d, the well-known stripe-shaped spin-contrast\cite{Hanke2017} with a periodicity of $2a$ is visible in all four different domains of the \mono phase. However, there are two different orientations of the stripes, as seen most clearly from the fast Fourier transforms (FFTs) selectively taken from surface areas in the four domains (insets in \autoref{fig2}c,d; note, that c is rotated by $45^\circ$ with respect to d). Note, that the angle between $q_{\rm a}$ and $q_{\rm b}$ differs from $90^\circ$ due to a residual lateral creep in these particular images, and that the $q_{\rm AFM}$ peaks appear rather blurry, because of the small image area. While the stripes in the red and yellow domain types are roughly oriented at a polar angle of $45^\circ$, they are oriented at $135^\circ$ in the blue and green domain types. Since the stripe shaped contrast can be assigned to the diagonal double-stripe (DDS) spin order of FeTe\cite{Hanke2017}, we can conclude that the DDSs have the same orientation in the red and yellow domains, but are rotated by about $90^\circ$ with respect to this orientation in the green and blue domains. This leads to two different types of domain boundaries; one where the orientation of the DDSs is not changing (A-type domain boundary between red and yellow, or blue and green domains, e.g. \autoref{fig2}c), and one where the DDSs are rotating by about $90^\circ$ (B-type domain boundary between yellow and green, or red and blue domains, e.g. \autoref{fig2}d). Note, that the DDS order changes its orientation at the B-type domain boundary not abruptly, but there is an area of a width of about $\SI{5}{nm}$ where the two orientations seem to coexist (\autoref{fig2}d). Nevertheless, the surface domain-types and the magnetic orientations are obviously interconnected and we presume, that the formation of the four domain types is due to the magneto-structural phase transition in the Fe$_{1.08}$Te sample.

\begin{figure}[H]
	\includegraphics[width = \columnwidth]{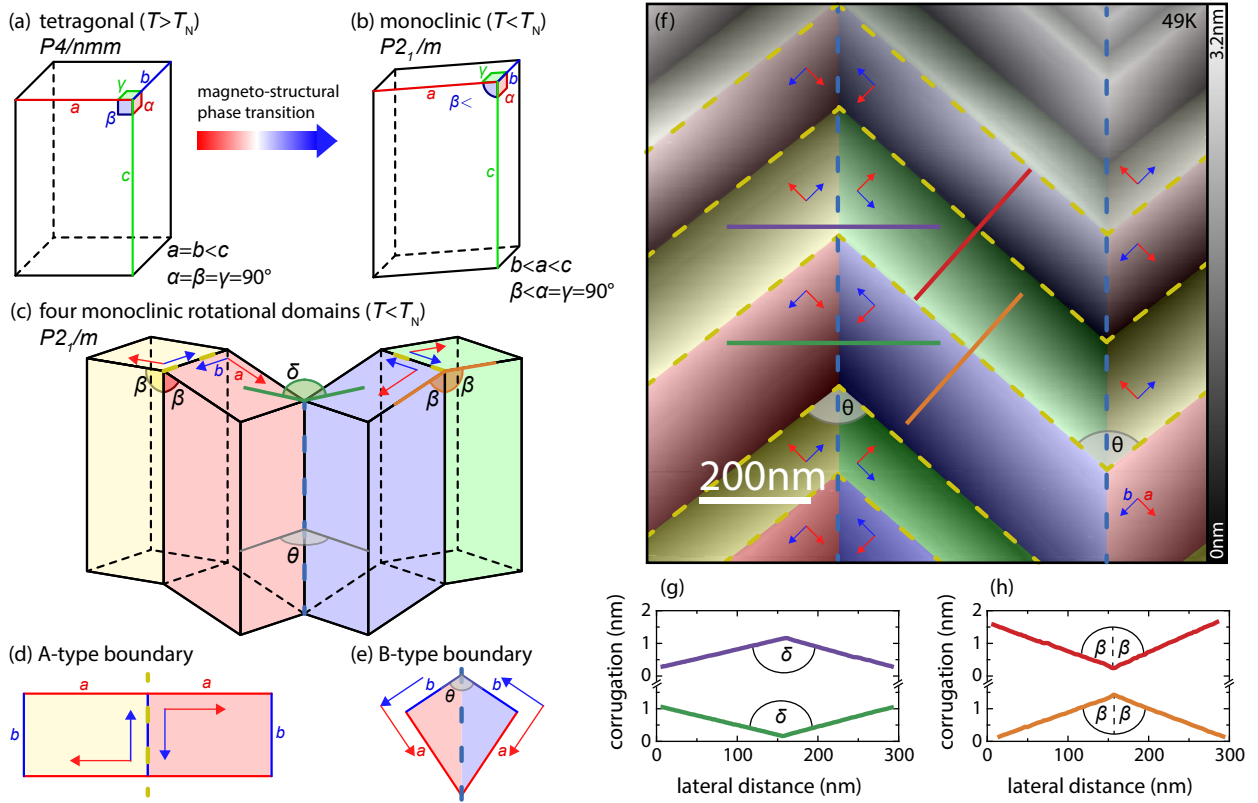} 
 	\caption{\label{fig3} \textbf{Structural models of \FeTe in the \tetra and \mono phases.} \textbf{a}, \textbf{b}, Simplified unit cell models for the \tetra phase (\textbf{a}) and for the \mono phase (\textbf{b}) with the indicated directions of the $a$, $b$, and $c$ lattice vectors and the corresponding angles. \textbf{c}, Illustration of the proposed structural model for the \mono phase of Fe$_{1.08}$Te with four rotational domains composed of unit cells which are rotated by $0^\circ$ (yellow), $180^\circ$ (red), $\theta$ (blue), and $\theta+180^\circ$ (green). Vectors in the direction of $a$ (red), and $b$ (blue), as well as the definition of the angles $\beta$ and $\delta$ are indicated. \textbf{d}, \textbf{e}, Top views of the A-type (\textbf{d}) and B-type (\textbf{e}) structural domain boundaries (same domain colors, vectors and angles as in \textbf{c}). \textbf{f}, Constant-current STM image of the surface of Fe$_{1.08}$Te below $T_{\rm N}$ overlayed with a color map marking the four rotational structural domains with the same colors as in \textbf{c} ($V_{\rm s}=\SI{300}{\milli\volt}$, $I_{\rm t}=\SI{40}{\pico\ampere}$). The dashed yellow and blue lines mark the A- and B-type boundaries, respectively. The angle $\theta$ between A-type boundaries is indicated. \textbf{g}, \textbf{h}, Profiles perpendicular across two B-type (\textbf{g}) and two A-type (\textbf{h}) boundaries. The profiles were taken along the lines which are marked with the identical colors in \textbf{f}. The extracted angles $\delta$ and $\beta$ are indicated. }
\end{figure}

This conclusion is further substantiated by a structural model described in the following. Since the DDSs are known to be oriented along the lattice vector $b$, we can determine the $a$ and $b$ directions within the four different domains, as shown by the arrows in \autoref{fig2}c,d. In the \tetra phase the four rotational domains, which can be generated by $90^\circ$ rotations of the tetragonal lattice around the $c$ axis, are degenerate (\autoref{fig3}a). However, this is not the case for the \mono phase (\autoref{fig3}a). Here, the four $90^\circ$ rotations of the monoclinic lattice result in four distinguishable orientations which are shown in \autoref{fig3}c. Each of the four orientations has a distinct normal vector of the $(a,b)$ plane. We therefore tentatively conclude, that the four domains observed at the surface in the STM images are due to four rotational structural domains of the crystal, where in each domain the lattice is given by one of the four orientations shown in \autoref{fig3}c. This is finally corroborated by a comparison of the expected and measured angles between the local surfaces in each of the four domains. Line profiles taken perpendicularly across B-type (\autoref{fig3}f,g) and A-type (\autoref{fig3}f,h) boundaries, respectively, reveal angles of $\delta_\text{STM}=178.93^\circ\pm0.03^\circ$ and $2\beta_\text{STM}=178.38^\circ\pm0.08^\circ$ (averages of 15 line profiles). The according angles expected within the structural model of \autoref{fig3}c are $2\beta=178.424^\circ\pm0.0006^\circ$ known from neutron powder-diffraction of Fe$_{1.068}$Te \cite{Li2009a}, and \mbox{$\delta=-(\cos^{-1}(\cos^2\beta\cdot\cos\theta+\sin^2\beta)-180^\circ)$} which depends on the rotation $\theta$ between the red and blue domain. $\beta_\text{STM}$ is in excellent agreement with the neutron diffraction data. For the determination of $\delta$, we need to consider that $\theta$ is not exactly $90^\circ$ because of the difference in the $a$ and $b$ lattice constants (\autoref{fig3}e) which will also induce strain in the B-type boundary. While it is a complex problem to predict the resulting $\theta$ from the model, we can still measure $\theta_\text{STM}$ from the experimental data as the angle between A-type boundaries (\autoref{fig3}f), which results in $\theta_\text{STM}=104.0^\circ\pm1.5^\circ$ (average from 15 measurements). Note that these measurements have been done using similar images as that shown in \autoref{fig3}f, but taken at \SI{32}{\kelvin} where the STM is most stable and the length calibration is most accurate. Indeed $\theta$ is considerably larger than $90^\circ$. The resulting calculated $\delta=178.76^\circ\pm0.01^\circ$ (based on $\beta$ and $\theta_\text{STM}$) is consistent with the measured value given above. We can therefore conclude, that the observed domains are indeed the four rotational structural domains illustrated in \autoref{fig3}c.

\begin{figure}[H]

	\includegraphics[width = 0.7\columnwidth]{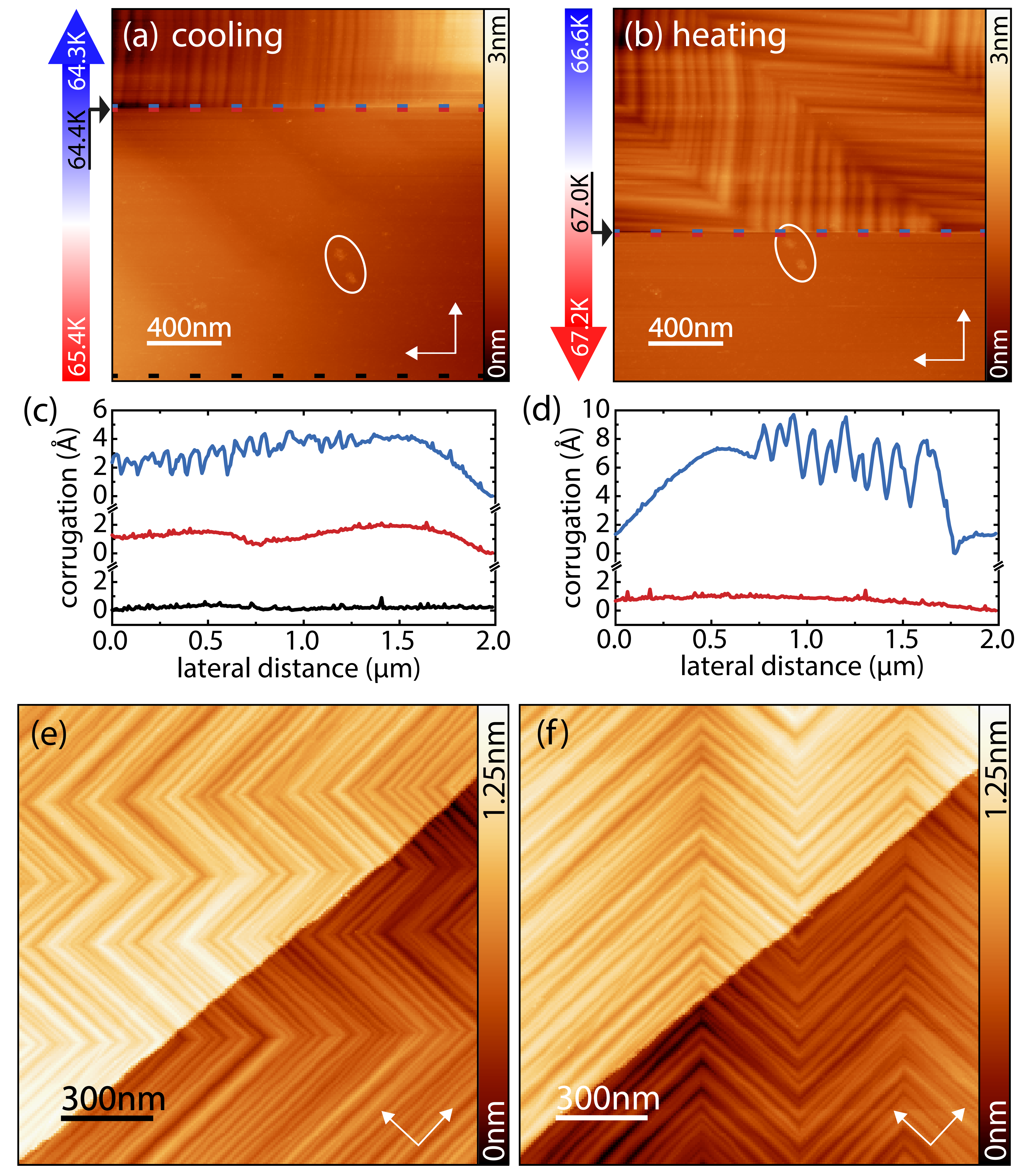} 
 	\caption{\label{fig4} \textbf{Domain imaging across the magneto-structural phase transition of Fe$_{1.08}$Te.}\\ \textbf{a}, Constant-current STM image of Fe$_{1.08}$Te taken while decreasing the temperature with a rate of $\SI{0.1}{\kelvin\per\minute}$ across the magneto-structural phase transition ($V_{\rm s}=\SI{300}{\milli\volt}$, $I_{\rm t}=\SI{40}{\pico\ampere}$). The image is scanned from bottom to top, line by line in the horizontal direction, resulting in a linearly increasing temperature as given by the temperature scale on the left. Each of the 256 lines takes a time of $\approx\SI{2.66}{\second}$. The phase transition from the \tetra (bottom) to the \mono (top) phase occurs at $T=\SI{64.4}{\kelvin}$. \textbf{b}, Same as \textbf{a}, but during increasing the temperature across the magneto-structural phase transition from the top to the bottom of the image. The scanned area is the same as in \textbf{a} as indicated by the defects used as a marker (see circles in \textbf{a} and \textbf{b}). The phase transition from \mono (top) to \tetra (bottom) phase occurs at $T=\SI{67.0}{\kelvin}$. \textbf{c}, \textbf{d}, Height profiles in the \mono (blue) and \tetra (red, black) phases for the cooling (\textbf{c}) and the heating (\textbf{d}) cycles, taken along the dashed lines marked with the corresponding color in \textbf{a} and \textbf{b}, respectively. \textbf{e}, \textbf{f}, Constant-current STM images of the identical surface area in the \mono phase of Fe$_{1.08}$Te, containing a step edge ($V_{\rm s}=\SI{300}{\milli\volt}$, $I_{\rm t}=\SI{40}{\pico\ampere}$). Inbetween the images, the sample was heated into the \tetra phase and cooled back into the \mono phase.}
 \end{figure}

The identity between structural and magnetic domains shown above enables measurements of the magneto-structural domains during cooling or heating across the magneto-structural phase transition without the need of spin-resolution in STM, as will be shown in the following. \autoref{fig4}a illustrates an STM image of the surface of the Fe$_{1.08}$Te sample which has been taken while cooling the sample from a temperature slightly above to a temperature slightly below the critical temperature using a constant cooling rate. Since the cooling rate ($\SI{0.1}{\kelvin\per\minute}$) is very slow as compared to the scan rate ($\SI{2.66}{\second}$ per line) the temperature decreases by less than $\SI{5}{\milli\kelvin}$ in each scan line from bottom to top. In the \tetra phase (bottom of \autoref{fig4}a and black line profile in \autoref{fig4}c) the surface is atomically flat. When approaching the phase transition, the surface starts to show some waviness in the diagonal direction with a characteristic length scale of $\approx\SI{1}{\mu\meter}$ (center of \autoref{fig4}a and red line profile in \autoref{fig4}c). Finally, at $T=\SI{64.4}{\kelvin}$, the imaged sample area passes over into the \mono phase as indicated by the sudden appearance of the strong corrugation in the blue line profile in \autoref{fig4}c, which is due to the magneto-structural domains (top of \autoref{fig4}a). This transition happens within a single scan line, i.e. in a temperature interval on the order of only $\SI{5}{\milli\kelvin}$. However, due to the slow scanning, we cannot determine the time it takes for the whole image area to transition into the \mono phase. A similar experiment is shown in \autoref{fig4}b, but now during heating of the sample across the phase transition while scanning from top to bottom. Note, that the area is exactly the same as in \autoref{fig4}a as indicated by the defects used as markers (see circles). The corresponding line profiles directly before and after the phase transition are given in \autoref{fig4}d as blue and red lines, respectively. Again, the transition happens very abruptly in one scan line. However, now, the transition temperature is considerably higher ($T=\SI{67.0}{\kelvin}$) as for the cooling cycle. Obviously, this thermal hysteresis ($\Delta T \approx \SI{2.5}{\kelvin}$) for the phase transition of the small imaged surface area is of a similar size as that of the macroscopic sample determined from the susceptibility measurements given above (\autoref{fig1}a). Interestingly, the long scale waviness of the surface in the \tetra phase, which was found as a precursor of the phase transition for the cooling cycle, is not visible after the transition from the \mono into the \tetra phase in the heating cycle (bottom of \autoref{fig4}b and red line profile in \autoref{fig4}d). This precursor is probably a strongly strained \tetra phase, which is not present in the heating cycle because of the increased temperature due to thermal hysteresis.

Interesting questions are, whether the arrangement of the magneto-structural domains is affected by the surface, and whether it changes after consecutive phase transitions. In order to investigate these questions, we took an STM image of a sample area which is crossed by a step edge of unit cell height in the \mono phase, heated into the \tetra phase, cooled back into the \mono phase, and imaged the same sample area (\autoref{fig4}e,f). Obviously, the arrangement of the magneto-structural domains changed drastically after cycling the sample once through the phase transition. While the chevron pattern of the domain structure is running vertically in \autoref{fig4}e, it runs horizontally in \autoref{fig4}f, indicating a $90^\circ$ rotation of the B-type domain boundaries. Moreover, the length of the domains has changed, as visible from the increased separation of the B-type domain boundaries in \autoref{fig4}f as compared to \autoref{fig4}e. Also, the domain width is considerably decreased in the images \autoref{fig4}e,f which were taken after cycling the sample slowly through the phase transition, as compared to the more virgin sample in \autoref{fig1}b and \autoref{fig3}f. Although this has not been investigated systematically, it indicates that slowly cooling the sample through the phase transition decreases the domain width. Finally, there is no obvious effect of the step edge on the domain arrangement, indicating that the surface and their defects have a negligible impact on the magneto-structural domain configuration, which is a bulk phenomenon.

\begin{figure}[H]

	\includegraphics[width = 0.6\columnwidth]{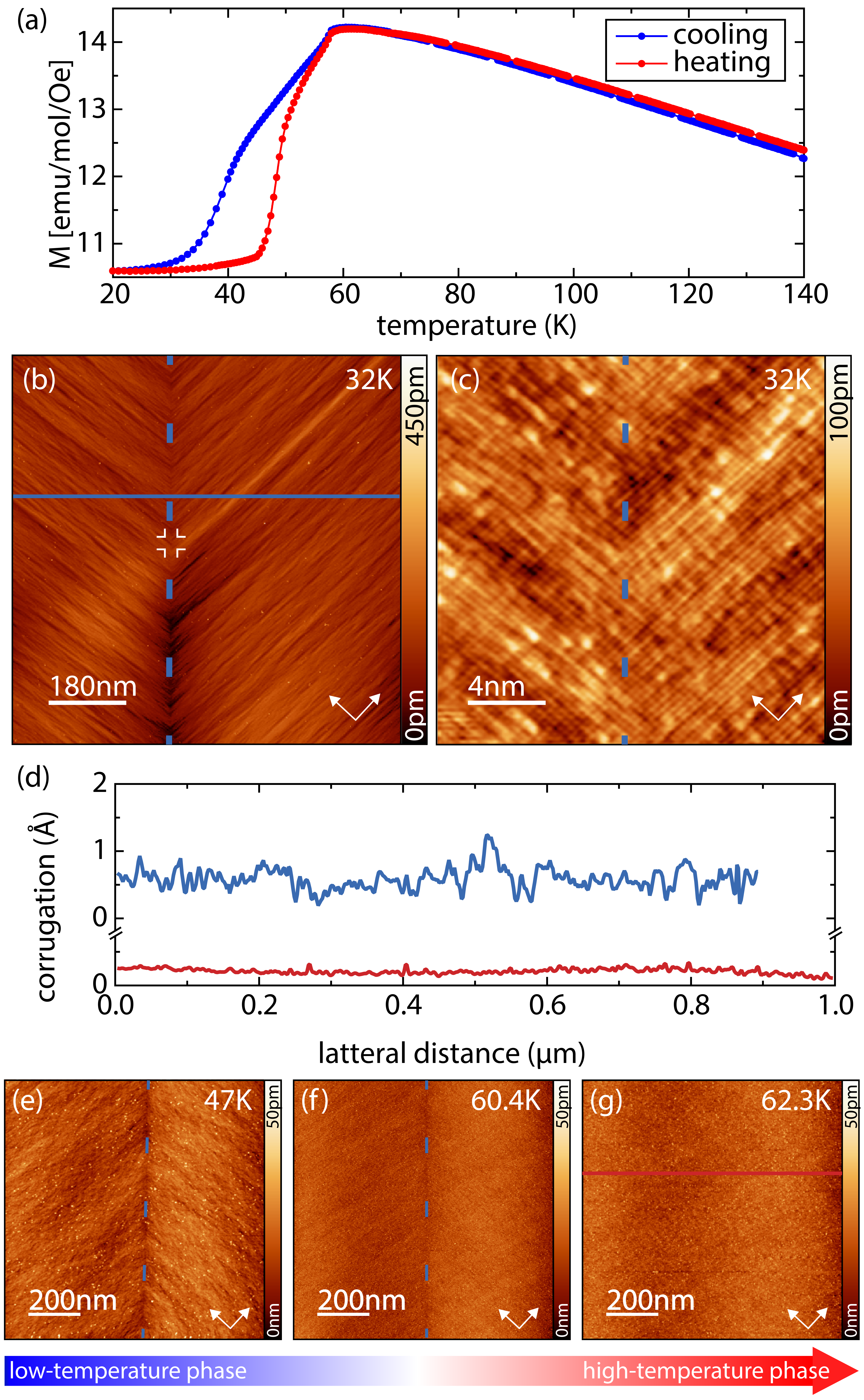} 
 	\caption{\label{fig5}\textbf{Domain imaging across the magnetic phase transition of Fe$_{1.12}$Te.}\\ \textbf{a}, Magnetic susceptibility of Fe$_{1.12}$Te measured in a field of \SI{0.1}{\tesla} as a function of temperature for the cooling (blue) and heating (red) cycle. There is a rather broad phase transition starting at \TN$\approx\SI{59}{\kelvin}$. \textbf{b}, Overview constant-current STM image of the sample surface in the low temperature phase taken at $T=\SI{32}{\kelvin}$ ($V_{\rm s}=\SI{300}{\milli\volt}$, $I_{\rm t}=\SI{40}{\pico\ampere}$). A B-type boundary is marked by the dashed vertical line. \textbf{c}, SP-STM image taken at the same temperature close to the B-type boundary showing the diagonal double-stripe spin orders in the two domains which are rotated by $90^\circ$ with respect to each other ($V_{\rm s}=\SI{-20}{\milli\volt}$, $I_{\rm t}=\SI{1}{\nano\ampere}$). \textbf{d} Blue and red height profiles perpendicular to the B-type boundary taken along the horizontal lines marked in \textbf{b} and \textbf{g}, respectively. \textbf{e} to \textbf{g}, Constant-current STM images of an identical surface area taken across the phase transition at the indicated temperatures $T=\SI{47}{\kelvin}$ (\textbf{e}), $T=\SI{60.4}{\kelvin}$ (\textbf{f}), and $T=\SI{62.3}{\kelvin}$ (\textbf{g}). A B-type domain boundary is marked by the dashed vertical lines in \textbf{e} and \textbf{f}, but is now longer visible in \textbf{g} ($V_{\rm s}=\SI{300}{\milli\volt}$, $I_{\rm t}=\SI{40}{\pico\ampere}$).}
 \end{figure}

Lastly, we investigate the effect of an increased excess Fe content $y$ by imaging the domains at the surface of a Fe$_{1.12}$Te sample across the magnetic phase transition (\autoref{fig5}). The temperature dependence of the magnetic susceptibility of this sample (\autoref{fig5}a) reveals a much broader phase transition starting at a lower \TN$\approx\SI{59}{\kelvin}$, as compared to the Fe$_{1.08}$Te sample (cf. \autoref{fig1}a). Moreover, there is an additional shoulder at lower temperature showing a strong thermal hysteresis in the cooling and heating cycles of about $\Delta T \approx \SI{10}{\kelvin}$. The characteristic shape of the susceptibility of the Fe$_{1.12}$Te sample was previously interpreted by a two-step structural phase transition\cite{Koz2013}: a second order transition from the tetragonal into an intermediate orthorhombic phase ($T\approx\SI{59}{\kelvin}$), followed by a first order transition into the monoclinic structure, as indicated by the shoulder ($T\approx\SI{40}{\kelvin}$). The strong thermal hysteresis of this shoulder indicates a considerable sluggishness of the latter phase transition which can be ascribed to a strong competition between orthorhombic and monoclinic phases. Therefore, it was proposed that the low temperature phase consists of a mixture of orthorhombic and monoclinic phases with metastable states that can persist over long time periods\cite{Koz2013}.

STM images in the low-temperature phase (\autoref{fig5}b) reveal a much less ordered structure as compared to the chevron pattern of the Fe$_{1.08}$Te sample (cf. \autoref{fig4}f). In atomically resolved SP-STM images, we can still observe a stripe-shaped spin-contrast (\autoref{fig5}c) with a clear B-type boundary marked by the dashed vertical line that separates domains where the stripes run from top left to bottom right (left of the boundary) and from bottom left to top right (right of the boundary). This B-type boundary is also well defined on a large length scale (dashed vertical line in \autoref{fig5}b). However, there is no clear formation of sharp and straight A-type boundaries as in the case of the Fe$_{1.08}$Te sample. Merely, an increased irregular surface roughness is visible in the direction perpendicular to the spin-contrast stripes (\autoref{fig5}d) which leads to the stripy appearance of the STM image in \autoref{fig5}b. When the sample is heated into the phase transition regime (\autoref{fig5}e,f) the stripy contrast and the B-type boundary very gradually decrease in visibility, until the sample surface finally gets atomically flat in the \tetra phase (\autoref{fig5}d,g). Lastly, we discuss the interpretation of these experimental findings with regard to the question whether there are indications for the proposed mixture of orthorhombic and monoclinic phases. Note, that the large scale images (\autoref{fig5}b,e,f), in particular those taken at $\SI{32}{\kelvin}$ and $\SI{47}{\kelvin}$, do not show any signatures of a long scale phase separation, at least not on the imaged area of $\SI{1}{\mu\meter}^2$. A pure \mono phase can be excluded, as there is no clear formation of chevron patterned domains which indicated the \mono phase in the Fe$_{1.08}$Te sample (cf. \autoref{fig4}f). Also, as revealed by the stripy appearance of the STM images, there is no pure orthorhombic phase, which would merely have B-type boundaries, but still some tendency towards \mono phase formation. We propose two possible scenarios: Either the FeTe layers of unit-cell thickness, which the sample is composed of, alternate between orthorhombic and \mono phases, with an orthorhombic surface layer, such that the surface has ripples from the burried monoclinic layers. Or, each layer contains a mixture of orthorhombic and monoclinic unit cells which alternate laterally on a length scale of roughly $\SI{5}{\nano\meter}$, which would then be the reason for the stripy appearance of the STM images (\autoref{fig5}c).

To summarize, we have presented the first simultaneous experimental investigation of the spatially resolved structure and magnetism in Fe$_{1+y}$Te which is continuously driven through the magnetic phase transition. For low excess iron content, the structural and magnetic domains are identical, as expected from a simultaneous magnetic and structural transition. For an excess iron content in the intermediate regime ($y=0.12$), we found evidence for a mixing of orthorhombic and monoclinic phases, which are either separated in the different layers, or in stripes of a couple of nanometer width. Our results show how the structural and the magnetic domains and phase transitions in the parent compound of a prototypical Fe-based superconductor are intertwined on the atomic scale. They reveal subtle effects as a leaking of the DDS spin order across the B-type domain boundaries (\autoref{fig2}d), a drastic shrinking of the domain size by slow phase transitions (\autoref{fig4}e), and a qualitative change induced by a slight increase in the excess Fe content. While the used technique is surface sensitive, the observed domain structure is most likely reflecting the structure in the bulk of the material, as we have seen that surface defects do not show any influence on the structural domains (\autoref{fig4}e,f). Our methodology therefore represents a pathway to investigate how magnetic and structural order can be established or suppressed by an appropriate treatment of the samples. As suppression of the magnetic order in Fe-based materials usually leads to the promotion of superconductivity, a similar study for the related superconducting compound Fe$_{1+y}$Se$_x$Te$_{1-x}$ will give profound insights into the question how superconductivity emerges from the magnetically ordered phase.

\begin{methods}

\subsection*{Spin-resolved scanning tunneling microscopy}
Temperature-dependent SP-STM measurements were performed using a home-built variable temperature STM \cite{Eich2014, Warmuth2016} located in a commercially available ultra-high vacuum (UHV) system, in which the samples have been treated prior to the presented measurements. The base pressure of the STM chamber was kept below $1\cdot10^{-10}\,$\si{\milli\bar} at all times. The tip and sample temperatures were controlled by a liquid Helium (He) flow cryostat covering temperatures from \RT down to \SI{30}{\kelvin}. The used chromium (Cr) bulk tip was electrochemically etched {\it ex-situ} and treated {\it in-situ} via field emission against a W(110) substrate before measuring\cite{Manna2017}. In order to finally achieve a stable spin contrast, the tip ``apex'' was repeatedly changed by applying rather high voltages of $1-2\,$\si{\volt} and high currents of $1-2\,$\si{\nano\ampere} while scanning the \FeTe surface, preferably when crossing a step edge. SP-STM images were recorded in constant-current mode at a temperature $T$, using a tunneling current ($I_{\rm t}$) and sample bias ($V_{\rm s}$), as stated individually for each image. The crystallographic orientation of the surface, known from atomically resolved STM images, is indicated by arrows in the constant-current STM images.

\subsection*{Magnetic susceptibility and X-ray diffraction measurements}
Magnetic susceptibility and single crystal X-ray diffraction (SC-XRD) measurements were performed on Fe$_{1.08}$Te and Fe$_{1.12}$Te crystal pieces originating from the same region of the \FeTe boule as the crystals investigated by STM. Thereby, we took special care to ensure the same composition, i.e. an identical excess Fe content $y$, of the samples investigated by all used methods. The magnetic susceptibility was measured on a Quantum Design physical property measurement system (PPMS) equipped with a vibrating sample magnetometer (VSM) as field cooling and heating (FCC-FCW) cycles in a magnetic field of $\SI{0.1}{\tesla}$.

\subsection*{Sample preparation}
\FeTe single crystal boules were grown by the Stockbarger-Bridgman method using iron pieces ($99.99\%$) and pieces of tellurium ingot ($99.999\%$). The starting materials with nominal compositions $y=0$ and $y=0.03$ each had a total mass of \SI{12}{\gram} and were loaded in quartz tubes (ID \SI{8}{\milli\meter}) with a conical tip and evacuated to a pressure of less than $4\cdot10^{-4}\,$\si{\milli\bar} and then sealed. The tube was then sealed in a larger quartz tube (ID $12\,$mm). A prereaction was performed at \SI{1000}{\celsius} for \SI{24}{\hour} to ensure homogeneity of the starting material. The tube was then inserted in a vertical tube furnace maintained at a fixed temperature of \SI{965}{\celsius} and translated out of the hot zone at a rate of $\SI{2}{\milli\meter\per\hour}$. At the melting point of \SI{914}{\celsius} the thermal gradient was measured to be \SI{25}{\celsius\per\centi\meter}. The obtained crystal boule could be cleaved across the entire diameter using a razor blade. Single crystal slabs with composition Fe$_{1.08}$Te and Fe$_{1.12}$Te as determined by inductively coupled plasma optical emission spectrometry (ICP-OES), were selected for further characterization by STM.

Two \FeTe crystals were cleaved several times at ambient conditions by the scotch tape method before introducing them into the UHV chamber. Subsequently, the \FeTe crystals were cleaved with the same method under UHV conditions ($p<10^{-10}\,$\si{\milli\bar}). This results in a clean, uncontaminated, and atomically flat surface\cite{Arnold:ArXiv2017}. The cleaving procedure splits the sample along the weakly bound van der Waals gaps between the FeTe layers. This way, STM measurements were always performed on the topmost Te layer. After cleaving under UHV conditions, the \FeTe samples exhibit single Fe atoms at the surface. These are observed as small circular protrusions in the STM images\cite{He2011} and stem from the excess Fe ($y\neq0$) located within the van der Waals gaps of bulk \FeTe crystals. The Fe atoms can be removed by a mild annealing process. Annealing the cleaved \FeTe crystal at \SI{430}{\kelvin} for \SI{30}{\minute} results in a clean \FeTe surface\cite{Arnold:ArXiv2017}. The surface corrugation and transition temperature \TN of \FeTe are not affected by this mild annealing procedure as verified by the comparison of Fe$_{1.08}$Te crystals before and after the process. This indicates that the intrinsic excess Fe amount $y$ is not affected by the annealing procedure as well. 
Merely the topmost excess Fe atoms are removed, thereby improving the quality of the atomically resolved SP-STM images. All presented STM measurements on Fe$_{1.08}$Te were done on annealed samples except for the data shown in \autoref{fig4}a and b. The investigation of the Fe$_{1.12}$Te single crystal was carried out on non-annealed samples.

\end{methods}

\newpage
\noindent \textbf{References}

\begin{thebibliography}{10}
\expandafter\ifx\csname url\endcsname\relax
  \def\url#1{\texttt{#1}}\fi
\expandafter\ifx\csname urlprefix\endcsname\relax\def\urlprefix{URL }\fi
\providecommand{\bibinfo}[2]{#2}
\providecommand{\eprint}[2][]{\url{#2}}

\bibitem{Paglione2010a}
\bibinfo{author}{Paglione, J.} \& \bibinfo{author}{Greene, R.~L.}
\newblock \bibinfo{title}{{High-temperature superconductivity in iron-based
  materials}}.
\newblock \emph{\bibinfo{journal}{Nat. Phys.}} \textbf{\bibinfo{volume}{6}},
  \bibinfo{pages}{645--658} (\bibinfo{year}{2010}).
\newblock \urlprefix\url{http://www.nature.com/doifinder/10.1038/nphys1759}.
\newblock \eprint{1006.4618}.

\bibitem{Stewart2011}
\bibinfo{author}{Stewart, G.~R.}
\newblock \bibinfo{title}{Superconductivity in iron compounds}.
\newblock \emph{\bibinfo{journal}{Rev. Mod. Phys.}}
  \textbf{\bibinfo{volume}{83}}, \bibinfo{pages}{1589--1652}
  (\bibinfo{year}{2011}).
\newblock \urlprefix\url{https://link.aps.org/doi/10.1103/RevModPhys.83.1589}.

\bibitem{Hoffman2011}
\bibinfo{author}{Hoffman, J.~E.}
\newblock \bibinfo{title}{Spectroscopic scanning tunneling microscopy insights
  into {Fe}-based superconductors}.
\newblock \emph{\bibinfo{journal}{Reports on Progress in Physics}}
  \textbf{\bibinfo{volume}{74}}, \bibinfo{pages}{124513}
  (\bibinfo{year}{2011}).
\newblock \urlprefix\url{http://stacks.iop.org/0034-4885/74/i=12/a=124513}.

\bibitem{Dai2015}
\bibinfo{author}{Dai, P.}
\newblock \bibinfo{title}{{Antiferromagnetic order and spin dynamics in
  iron-based superconductors}}.
\newblock \emph{\bibinfo{journal}{Rev. Mod. Phys.}}
  \textbf{\bibinfo{volume}{87}}, \bibinfo{pages}{855--896}
  (\bibinfo{year}{2015}).
\newblock \urlprefix\url{https://link.aps.org/doi/10.1103/RevModPhys.87.855}.
\newblock \eprint{1503.02340}.

\bibitem{He2011}
\bibinfo{author}{He, X.} \emph{et~al.}
\newblock \bibinfo{title}{Nanoscale chemical phase separation in
  {Fe}{Te}$_{0.55}${Se}$_{0.45}$ as seen via scanning tunneling spectroscopy}.
\newblock \emph{\bibinfo{journal}{Phys. Rev. B}} \textbf{\bibinfo{volume}{83}},
  \bibinfo{pages}{220502} (\bibinfo{year}{2011}).
\newblock \urlprefix\url{https://link.aps.org/doi/10.1103/PhysRevB.83.220502}.

\bibitem{Singh2013}
\bibinfo{author}{Singh, U.~R.} \emph{et~al.}
\newblock \bibinfo{title}{Spatial inhomogeneity of the superconducting gap and
  order parameter in {Fe}{Se}$_{0.4}${Te}$_{0.6}$}.
\newblock \emph{\bibinfo{journal}{Phys. Rev. B}} \textbf{\bibinfo{volume}{88}},
  \bibinfo{pages}{155124} (\bibinfo{year}{2013}).
\newblock \urlprefix\url{https://link.aps.org/doi/10.1103/PhysRevB.88.155124}.

\bibitem{Kamlapure2017}
\bibinfo{author}{Kamlapure, A.} \emph{et~al.}
\newblock \bibinfo{title}{{Spatial variation of the two-fold anisotropic
  superconducting gap in a monolayer of {Fe}{Se}$_{0.5}${Te}$_{0.5}$ on a
  topological insulator}}.
\newblock \emph{\bibinfo{journal}{Phys. Rev. B}} \textbf{\bibinfo{volume}{95}},
  \bibinfo{pages}{104509} (\bibinfo{year}{2017}).
\newblock \urlprefix\url{https://link.aps.org/doi/10.1103/PhysRevB.95.104509}.

\bibitem{Enayat2014a}
\bibinfo{author}{Enayat, M.} \emph{et~al.}
\newblock \bibinfo{title}{Real-space imaging of the atomic-scale magnetic
  structure of {Fe}$_{1+y}${Te}}.
\newblock \emph{\bibinfo{journal}{Science (80-. ).}}
  \textbf{\bibinfo{volume}{345}}, \bibinfo{pages}{653--656}
  (\bibinfo{year}{2014}).
\newblock
  \urlprefix\url{http://www.sciencemag.org/cgi/doi/10.1126/science.1251682}.

\bibitem{Manna2017}
\bibinfo{author}{Manna, S.} \emph{et~al.}
\newblock \bibinfo{title}{{Interfacial superconductivity in a bi-collinear
  antiferromagnetically ordered FeTe monolayer on a topological insulator}}.
\newblock \emph{\bibinfo{journal}{Nat. Commun.}} \textbf{\bibinfo{volume}{8}},
  \bibinfo{pages}{14074} (\bibinfo{year}{2017}).
\newblock \urlprefix\url{http://www.nature.com/doifinder/10.1038/ncomms14074}.
\newblock \eprint{1606.03249}.

\bibitem{Hanke2017}
\bibinfo{author}{H{\"{a}}nke, T.} \emph{et~al.}
\newblock \bibinfo{title}{{Reorientation of the diagonal double-stripe spin
  structure at Fe$_{1+y}$Te bulk and thin-film surfaces}}.
\newblock \emph{\bibinfo{journal}{Nat. Commun.}} \textbf{\bibinfo{volume}{8}},
  \bibinfo{pages}{13939} (\bibinfo{year}{2017}).
\newblock \urlprefix\url{http://www.nature.com/doifinder/10.1038/ncomms13939}.

\bibitem{Koz2013}
\bibinfo{author}{Koz, C.}, \bibinfo{author}{R{\"{o}}{\ss}ler, S.},
  \bibinfo{author}{Tsirlin, A.~A.}, \bibinfo{author}{Wirth, S.} \&
  \bibinfo{author}{Schwarz, U.}
\newblock \bibinfo{title}{{Low-temperature phase diagram of Fe$_{1+y}$Te
  studied using x-ray diffraction}}.
\newblock \emph{\bibinfo{journal}{Phys. Rev. B}} \textbf{\bibinfo{volume}{88}},
  \bibinfo{pages}{094509} (\bibinfo{year}{2013}).
\newblock \urlprefix\url{https://link.aps.org/doi/10.1103/PhysRevB.88.094509}.

\bibitem{Mizuguchi2010}
\bibinfo{author}{Mizuguchi, Y.} \& \bibinfo{author}{Takano, Y.}
\newblock \bibinfo{title}{{Review of Fe chalcogenides as the simplest Fe-based
  superconductor}}.
\newblock \emph{\bibinfo{journal}{J. Phys. Soc. Japan}}
  \textbf{\bibinfo{volume}{79}}, \bibinfo{pages}{10--13}
  (\bibinfo{year}{2010}).
\newblock \urlprefix\url{http://journals.jps.jp/doi/abs/10.1143/JPSJ.79.102001
  http://arxiv.org/abs/1003.2696{\%}5Cnhttp://www.arxiv.org/pdf/1003.2696.pdf}.
\newblock \eprint{1003.2696}.

\bibitem{Subedi2008}
\bibinfo{author}{Subedi, A.}, \bibinfo{author}{Zhang, L.},
  \bibinfo{author}{Singh, D.~J.} \& \bibinfo{author}{Du, M.~H.}
\newblock \bibinfo{title}{{Density functional study of FeS, FeSe, and FeTe:
  Electronic structure, magnetism, phonons, and superconductivity}}.
\newblock \emph{\bibinfo{journal}{Phys. Rev. B - Condens. Matter Mater. Phys.}}
  \textbf{\bibinfo{volume}{78}}, \bibinfo{pages}{134514}
  (\bibinfo{year}{2008}).
\newblock \urlprefix\url{https://link.aps.org/doi/10.1103/PhysRevB.78.134514}.
\newblock \eprint{0807.4312}.

\bibitem{Zaliznyak2012}
\bibinfo{author}{Zaliznyak, I.~A.} \emph{et~al.}
\newblock \bibinfo{title}{{Continuous magnetic and structural phase transitions
  in Fe$_{1+y}$Te}}.
\newblock \emph{\bibinfo{journal}{Phys. Rev. B}} \textbf{\bibinfo{volume}{85}},
  \bibinfo{pages}{085105} (\bibinfo{year}{2012}).
\newblock \urlprefix\url{https://link.aps.org/doi/10.1103/PhysRevB.85.085105}.
\newblock \eprint{1108.5968}.

\bibitem{Bao2009b}
\bibinfo{author}{Bao, W.} \emph{et~al.}
\newblock \bibinfo{title}{Tunable ($\ensuremath{\delta}\ensuremath{\pi}$,
  $\ensuremath{\delta}\ensuremath{\pi}$)-type antiferromagnetic order in
  $\ensuremath{\alpha}$-{Fe}({Te},{Se}) superconductors}.
\newblock \emph{\bibinfo{journal}{Phys. Rev. Lett.}}
  \textbf{\bibinfo{volume}{102}}, \bibinfo{pages}{247001}
  (\bibinfo{year}{2009}).
\newblock
  \urlprefix\url{https://link.aps.org/doi/10.1103/PhysRevLett.102.247001}.

\bibitem{Rodriguez2011}
\bibinfo{author}{Rodriguez, E.~E.} \emph{et~al.}
\newblock \bibinfo{title}{Magnetic-crystallographic phase diagram of the
  superconducting parent compound {Fe}$_{1+x}${Te}}.
\newblock \emph{\bibinfo{journal}{Phys. Rev. B}} \textbf{\bibinfo{volume}{84}},
  \bibinfo{pages}{064403} (\bibinfo{year}{2011}).
\newblock \urlprefix\url{https://link.aps.org/doi/10.1103/PhysRevB.84.064403}.

\bibitem{Li2009a}
\bibinfo{author}{Li, S.} \emph{et~al.}
\newblock \bibinfo{title}{{First-order magnetic and structural phase
  transitions in Fe$_{1+y}$Se$_x$Te$_{1-x}$}}.
\newblock \emph{\bibinfo{journal}{Phys. Rev. B - Condens. Matter Mater. Phys.}}
  \textbf{\bibinfo{volume}{79}}, \bibinfo{pages}{054503}
  (\bibinfo{year}{2009}).
\newblock \urlprefix\url{http://link.aps.org/doi/10.1103/PhysRevB.79.054503
  https://link.aps.org/doi/10.1103/PhysRevB.79.054503}.

\bibitem{Eich2014}
\bibinfo{author}{Eich, A.} \emph{et~al.}
\newblock \bibinfo{title}{Intra- and interband electron scattering in a hybrid
  topological insulator: Bismuth bilayer on {Bi}$_2${Se}$_3$}.
\newblock \emph{\bibinfo{journal}{Phys. Rev. B - Condens. Matter Mater. Phys.}}
  \textbf{\bibinfo{volume}{90}}, \bibinfo{pages}{155414}
  (\bibinfo{year}{2014}).
\newblock \urlprefix\url{https://link.aps.org/doi/10.1103/PhysRevB.90.155414}.
\newblock \eprint{1409.7014}.

\bibitem{Warmuth2016}
\bibinfo{author}{Warmuth, J.} \emph{et~al.}
\newblock \bibinfo{title}{{Band-gap engineering by Bi intercalation of graphene
  on Ir(111)}}.
\newblock \emph{\bibinfo{journal}{Phys. Rev. B}} \textbf{\bibinfo{volume}{93}},
  \bibinfo{pages}{165437} (\bibinfo{year}{2016}).
\newblock \urlprefix\url{https://link.aps.org/doi/10.1103/PhysRevB.93.165437}.
\newblock \eprint{1603.08724}.

\bibitem{Arnold:ArXiv2017}
\bibinfo{author}{Arnold, F.} \emph{et~al.}
\newblock \bibinfo{title}{{Electronic structure of Fe$_{1.08}$Te bulk crystals
  and epitaxial FeTe thin films on Bi$_2$Te$_3$}}.
\newblock \emph{\bibinfo{journal}{arXiv:1711.07039v1 [cond-mat.str-el]}}
  (\bibinfo{year}{2017}).

\end{thebibliography}


\begin{addendum}
 \item J. Wi and P. H. acknowledge funding from the German Research Foundation via the DFG priority programme SPP1666 (grant nos. HO 5150/1-2 and WI 3097/2-2). R. W. acknowledges funding via the ERC Advanced Grand ASTONISH (number 338802). J. Wa. and R. W. acknowledge funding from the German Research Foundation via the SFB668. This work was supported by VILLUM FONDEN via the Centre of Excellence for Dirac Materials (Grant No. 11744). M. B. acknowledges funding from the Danish National Research Foundation under the grant DNRF93, Center for Materials Crystallography (CMC).

 \item[Author contributions] J. Wa., J. W., and M. B. designed the experiment. J. Wa. carried out and did the analysis of the SP-STM measurements. M. B. grew the samples, carried out and did the analysis of the magnetic susceptibility, SC-XRD, and ICP-OES measurements. J. Wa. prepared the figures. J. Wa. and J. W. wrote the manuscript with contributions from M. B. All authors contributed to the discussion and interpretation of the results as well as to the discussion of the manuscript.

 \item[Competing Interests] The authors declare that they have no competing financial interests.

 \item[Correspondence] Correspondence and requests for materials should be addressed to J. Warmuth.~(email: jwarmuth@physnet.uni-hamburg.de), or J. Wiebe~(email: jwiebe@physnet.uni-hamburg.de).
\end{addendum}

\newpage

\newpage

\newpage

\newpage

\newpage

\end{document}